\documentclass[12pt]{article}
\usepackage{amssymb}

\newcommand{\nc}{\newcommand}
\nc{\la}{\lambda} \nc{\alf}{\alpha}
\nc{\tht}{\theta}  \nc{\be}{\beta}  \nc{\eps}{\epsilon}
\nc{\ga}{\gamma}  \nc{\De}{\Delta}  \nc{\Ga}{\Gamma}  \nc{\vphi}{\varphi}
\nc{\de}{\delta} \nc{\si}{\sigma}  \nc{\ka}{\kappa}
\nc{\om}{\omega}  \nc{\Om}{\Omega}  
\nc{\qq}{\quad\quad}  \nc{\nf}{\infty}   \nc{\dl}{\mathop{\smash{\cal L}}}
\nc{\ra}{\rightarrow}  \nc{\ol}{\overline}  \nc{\und}{\underline}
\nc{\beq}{\begin{equation}}  \nc{\pt}{\partial}  \nc{\nin}{\noindent}
\nc{\eeq}{\end{equation}}
\nc{\beqa}{\begin{eqnarray}}  \nc{\dst}{\displaystyle}
\nc{\eeqa}{\end{eqnarray}} \nc{\nnb}{\nonumber}   \nc{\rl}{\rule{0mm}{4mm}}
\nc{\bs}{\backslash}        \nc{\mb}{\mathbb} \nc{\R}{\mb R}
\nc{\dg}{\dagger}   \nc{\ti}{\tilde}     \nc{\wh}{\widehat}  \nc{\wti}{\widetilde}
\nc{\px}{\Pi_x}  \nc{\py}{\Pi_{y}}

\newcounter{muni}
\newenvironment{remunerate}{\begin{list}{{\rm \arabic{muni}.}}
{\usecounter{muni}
\setlength{\leftmargin}{0pt}\setlength{\itemindent}{38pt}}}{\end{list}}

\nc{\brm}{\begin{remunerate}}   \nc{\erm}{\end{remunerate}}

\newtheorem{nth}{Proposition}  \nc{\stg}{\mathop{\smash{*}}}
\nc{\st}{\mathop{\smash{\delta}}}
\nc{\barr}{\begin{array}}   \nc{\earr}{\end{array}}   
\nc{\mtvb}{\mathversion{bold}}   \nc{\mtvn}{\mathversion{normal}}
\nc{\sta}{St\" ackel}

\begin{document}
\begin{titlepage}

\[  \]
\centerline{\Large\bf Neumann-like integrable models }

\vskip 2.0truecm
\centerline{\large\bf Galliano VALENT${}^{\;\dagger\; *}$}

\vskip 1.0truecm
\centerline{\large\bf Hamed Ben YAHIA${}^{\;\dagger}$}

\vskip 2.0truecm
\centerline{${}^{\dagger}$ \it Laboratoire de Physique Th\'eorique et des
Hautes Energies}
\centerline{\it CNRS, Unit\'e associ\'ee URA 280}
\centerline{\it 2 Place Jussieu, F-75251 Paris Cedex 05, France}
\nopagebreak

\vskip 0.5truecm
\centerline{${}^*$ \it D\'epartement de Math\'ematiques}
\centerline{\it UFR Sciences-Luminy}
\centerline{\it Case 901 163 Avenue de Luminy}
\centerline{\it 13258 Marseille Cedex 9, France}
\nopagebreak

\vskip 2.5truecm

\begin{abstract}
A countable class of integrable dynamical systems, with four dimensional 
phase space and conserved quantities in involution $(H_n,I_n)$ are exhibited. For 
$n=1$ we recover Neumann sytem on $T^*\,S^2$. All these systems are also integrable at the 
quantum level.
\end{abstract}

\end{titlepage}

\section{Introduction}
The classical problem of motion of a rigid body in an ideal fluid leads to one 
among the oldest integrable models : Clebsch dynamical system \cite{Cl}. Upon symplectic 
reduction it becomes Neumann celebrated integrable system \cite{Ne} with phase space 
$T^*S^2.$ Its Hamiltonian $H$ Poisson-commutes with an extra independent 
quadratic integral $I$, quadratic in the configuration space coordinates. Forgetting its physical 
interpretation and just considering it as a dynamical system, the possibility of 
finding generalizations of it was hopeless in view of a uniqueness theorem by 
Perelomov \cite{Pe1}. A close examination of the hypotheses under which this 
uniqueness result is obtained shows that some room is left for generalization if one does 
remain on $T^*\,S^2$. However the equations to be solved for these generalizations are somewhat 
involved and we were happy enough to get one using symbolic 
computation (Section 3). This example can be further generalized and gives rise to a countable 
family of integrable systems, which we show to be different from the family given earlier by 
Wojciechowski (Section 4). We conclude by proving that, using the so-called ``minimal" 
quantization, our class of models are also integrable at the quantum level (Section 5).

\section{Neumann integrable system}
This integrable system is defined from the Lie algebra ${\cal G}=e(3)$ with respect 
to the Poisson bracket defined by
\beq\label{cl1}
\{M_i,M_j\}=\eps_{ijk}\,M_k,\qq \{M_i,X_j\}=\eps_{ijk}\,X_k,\qq\{X_i,X_j\}=0,
\qq i,j,k=1,2,3.\eeq
The hamiltonian flow is given by
\beq\label{cl2}
\dot{M}_i=\{H,M_i\},\qq\quad \dot{X}_i=\{H,X_i\}.\eeq
It is easy to check that 
\[C_1=\sum_i\,X_i^2,\qq\&\qq C_2=\sum_i\,X_i\,M_i\]
are two Casimir functions. Considering them as constants, for instance $C_1=1$ 
and $C_2=0,$ we obtain the orbit of the 
co-adjoint representation of the group $G=E(3)$ which is the four-dimensional 
phase space $\Om=T^*\,S^2$ of the considered system. Neumann hamiltonian \cite{Pe2} 
can be taken as
\beq\label{cl3}\left\{\barr{c}
H=H^{(2)}+U,\\[4mm]
 H^{(2)}=\sum_i\,a_i\,M_i^2,\quad  U=a_2a_3\,X_1^2+a_3a_1\,X_2^2+a_1a_2\,X_3^2.
\earr\right.\eeq
Its integrability follows from the existence of the extra conserved quantity
\beq\label{cl4}\left\{\barr{c}
I=I^{(2)}+V,\\[4mm]
 I^{(2)}=\sum_i\,M_i^2,\quad V=(a_2+a_3)X_1^2+(a_3+a_1)X_2^2+(a_1+a_2)X_3^2,\earr
\right.\eeq
which Poisson-commutes with the Hamiltonian.

It is interesting to see how, given $H,$ one can construct $I$. Taking into account 
that $\{M_i,f\}=-\hat{L}_i\,f\equiv -\eps_{ijk}\,X_j\pt_k\,f$ one has
\beq\label{cl5}
\{H,I\}=\{H^{(2)},V\}-\{I^{(2)},U\}=2\sum_i\,M_i\, \wh{L}_i
\left(\rule{0mm}{4mm}U-a_i\,V\right).\eeq
So if we write $V=a\,X_1^2+b\,X_2^2+c\,X_3^2,$ the strict vanishing of the Poisson 
bracket requires
\beq\label{cl6}
b-c=-a_1(a_2-a_3),\qq c-a=-a_2(a_3-a_1),\qq a-b=-a_3(a_1-a_2).\eeq
Obviously these 3 relations add up to zero, so only two of them are independent 
and we get
\beq\label{cl7}
V=a\,C_1+(a_1a_3-a_2a_3)X_2^2+(a_1a_2-a_2a_3)X_3^2,\eeq
which displays the {\em uniqueness} of $V_1,$ up to the Casimir $C_1$. This uniqueness 
is stressed in proposition 1 of \cite{Pe1}. The choice $a=a_2a_3$ gives then the 
Neumann potential (\ref{cl4}).

\section{A first Neumann-like integrable system}
Our starting observation is simply that uniqueness is a result of the strong requirement 
of vanishing of the 3 terms appearing in (\ref{cl5}). This is certainly sufficient to 
get uniquely Neumann system, but it is not necessary. We could have, rather
\[\{H,I\}=\cdots(C_1-1)+\cdots C_2\]
and this is still conserved in $\Om.$ Under this weaker hypothesis we could hope for 
some Neumann-like integrable systems, with new potentials $U_2$ and $V_2.$ Indeed the 
equations to be integrated become
\beq\label{cl8}
\wh{L}_i\left(\rule{0mm}{4mm}U_2-a_i\,V_2\right)=\la_i(C_1-1)+\mu\,X_i,\eeq
where $(\la_i,\,\mu)$ are unknown functions of the $X_i.$ These equations are quite 
difficult to integrate in general, so we have been looking for a specific example where 
$U_2$ and $V_2$ are quartic polynomials \footnote{We were not able to find any 
solution with cubic polynomials.} and we have used Maple to solve for 
the equations. Quite surprisingly the solution, which is rather involved in the 
coordinates $X_i$, can be written in a rather simple form in terms of $U$ and $V$:
\beq\label{cl9}
U_2=U\,V\qq V_2=V^2-U.\eeq
Once this is observed, it is easy to give an analytic proof:

\begin{nth}The dynamical system $H_2=H^{(2)}+U_2$ and $I_2=I^{(2)}+V_2,$ with 
phase space $T^*\,S^2$, is integrable in Liouville sense.
\end{nth}

\nin{\bf Proof}: We start from
\beq\label{fin0}
 \wh{L}_1(U_2-a_1V_2)= (U-a_1V+a_1^2)\wh{L}_1V,\eeq
use the relation
\[U-a_1V+a_1^2=-(a_1-a_2)(a_3-a_1)X_1^2+a_1^2(1-C_1),\]
and
\[\wh{L}_1V=-2(a_2-a_3)X_2X_3,\qq \wh{L}_1U=a_1\,\wh{L}_1V,\]
which lead to
\beq\label{fin1}
\wh{L}_1(U_2-a_1V_2)= -2S(X)\,X_1+2a_1^2(a_2-a_3)X_2X_3(1-C_1),\eeq
with the totally symmetric function
\beq\label{fin2}
S(X)=(a_1-a_2)(a_2-a_3)(a_3-a_1)X_1X_2X_3.\eeq
Summing the various terms in (\ref{cl5}) we get
\beq\label{fin}\barr{l}
\{H_2,I_2\}=-4S(X)\,C_2+\\[4mm]
+4(1-C_1)\left[\rule{0mm}{4mm}
a_1^2(a_2-a_3)X_2X_3M_1+a_2^2(a_3-a_1)X_3X_1M_2+a_3^2(a_1-a_2)X_1X_2M_3\right],\earr
\eeq
which vanishes in $\Om.$ So this system is integrable . $\quad\Box$

\section{More Neumann-like integrable systems}
The previous result can be generalized to polynomials of even degree in the 
following way. Let us define the series $U_n$ and $V_n$ by the recurrence:
\beq\label{rec1}
\left\{\barr{l} U_1=U\\[4mm] V_1=V\earr\right.\qq\qq\qq\left\{\barr{l}
U_n=U\,V_{n-1}\\[4mm] V_n=V\,V_{n-1}-U_{n-1}\earr\right. \qq n\geq 2.
\eeq
Standard techniques give the following useful information on these polynomials:

\begin{nth} The explicit form of the polynomials is
\beq\label{rec2}
\left\{\barr{l}\dst 
U_n=\sum_{k=0}^{[(n-1)/2]}(-1)^k {n-1-k \choose k}U^{k+1}V^{n-1-2k},\\[6mm]\dst 
V_n=\sum_{k=0}^{[n/2]}(-1)^k {n-k \choose k}U^kV^{n-2k},\earr\right.
\eeq
and they verify the following partial differential equations:
\beq\label{rec3}
\pt_VU_n+U\pt_UV_n=0,\qq \pt_VV_n+V\pt_UV_n=\pt_UU_n,\qq n\geq 1.\eeq
\end{nth}

\nin{\bf Proof:}

\nin We first need to prove the three terms recurrence relation
\[V_{n+1}-VV_n+UV_{n-1}=0,\qq n\geq 2.\]
Using the following identity for the binomial coefficients
\[{n+1-k \choose k}-{n-k \choose k}={n-k \choose k-1}(1-\de_{k0}),\quad k\geq 0,\]
and the explicit form of $V_n$ one gets
\[V_{n+1}-VV_n=-\sum_{k=0}^{[(n-1)/2]}(-1)^k{n-1-k \choose k}U^{k+1}V^{n-1-2k}
=-UV_{n-1}.\]
The first partial differential equation follows from the identity
\[k{n-k \choose k}=(n+1-2k){n-k \choose k-1},\qq k\geq 1,\]
and the second one from
\[(n-2k){n-k \choose k}-(k+1){n-1-k \choose k}=(k+1){ n-1-k \choose k+1},\qq k\geq 0.\] 
In the analysis some care is required with the upper bounds of the summations.
$\quad\Box$

\nin We are now in position to prove:

\begin{nth} The dynamical systems $H_n=H^{(2)}+U_n$ and $I_n=I^{(2)}+V_n,$ with 
phase space $T^*\,S^2$, are integrable in Liouville sense.
\end{nth}

\nin{\bf Proof:}

\nin Using the recurrence relations for the polynomials $U_n$ and $V_n$ we have first
\[ \wh{L}_1(U_n-a_1V_n)=(U-a_1V)\wh{L}_1V_{n-1}+a_1\wh{L}_1U_{n-1}.\]
The generic relation
\[\wh{L}_1\,f=\left(\rule{0mm}{4mm}\pt_Vf+a_1\pt_Uf\right)\wh{L}_1V,\]
used in the previous equation gives for the right hand side
\[(U-a_1V)\pt_VV_{n-1}+a_1(U\pt_UV_{n-1}+\pt_VU_{n-1})
+a_1^2(\pt_U U_{n-1}-V\pt_U V_{n-1}).\]
The partial differential equations of proposition 2 give then
\beq\label{rec4}
\wh{L}_1(U_n-a_1V_n)= \partial_V V_{n-1}(U-a_1V+a_1^2)\wh{L}_1V=
\partial_V V_{n-1}\,\wh{L}_1(U_2-a_1V_2),
\eeq
and since $\partial_V V_{n-1}$ is fully symmetric we get
\beq\label{int1}
\{H_n,I_n\}=\partial_V V_{n-1}\{H_2,I_2\},\eeq
which proves the proposition. $\quad\Box$

Let us show that the family of potentials obtained here is indeed different from a family 
obtained by Wojciechowski in \cite{Wo}. This author has obtained a countable set of integrable potentials on $T^*\,S^n$ which we have to restrict to $n=2.$ There is no general formula for his potentials, but the simplest ones are given page 109 , which we will reproduce (we take of course $r=1$). The first three are \footnote{We discard Braden's potential $(\sum_k X_k^2/a_k)^{-1}$ since it is not a polynomial.}
\beqa
I & = & \sum_k a_k\,X_k^2, \label{w1} \\
I_2 & = & \sum_k a_k^2\,X_k^2-(\sum a_k\,X_k^2)^2,\label{w2}\\
\label{w3}
I_3 & = & \sum_k a_k^3\,X_k^2-2(\sum_j a_j\,X_j^2)(\sum_k a_k^2\,X_k^2)+(\sum_k a_k\,X_k^2)^3.\eeqa
Let us compare our potentials with these ones, beginning with $V.$ Using the constraint 
\beq\label{con}
C_1\equiv X_1^2+X_2^2+X_3^2=1,\eeq 
we can write
\[V=(a_2+a_3)X_1^2+(a_3+a_1)X_2^2+(a_1+a_2)X_3^2=\sum_k a_k-I\]
and, up to a constant, it coincides with the potential (\ref{w1}) given by W. This corresponds to Neumann on $T^*\,S^2.$

Then our potential $V_2$  is given by
\[V_2=-U+V^2=-(a_2a_3\,X_1^2+a_3a_1\,X_2^2+a_1a_2\,X_3^2)+(\sum_k a_k-\sum_k a_k\,X_k^2)^2,\]
to be compared with the quartic potential $I_2$. One can check the relation
\[V_2+I_2=(a_2a_3-a_1^2)X_1^2+(a_3a_1-a_2^2)X_2^2+(a_1a_2-a_3^2)X_3^2+a_1^2+a_2^2+a_3^2,\]
which shows that the quartic terms in $V_2$ and $I_2$ are just opposite in sign but that their quadratic content is {\em different} even using the constraint (\ref{con}).
So our potential $V_2$ is definitely different from the potential $I_2$ given by Wojciechowski.

Let us now consider our potential $V_3$ against $I_3.$ From the recurrence given in our article we have
\[\barr{l}
V_3=-2UV+V^3=\\[4mm] 
\hspace{1cm}=-2(\sum_k a_k-\sum_k a_k\,X_k^2)(a_2a_3\,X_1^2+a_3a_1\,X_2^2+a_1a_2\,X_3^2)+(\sum_k a_k-\sum_k a_k\,X_k^2)^3,\earr\]
and upon expanding, using the constraint and with some algebra we get
\beq\label{w4}
\barr{l}  \hspace{-3cm}
V_3+I_3=(a_1-a_2)(a_1-a_3)(3a_1+2a_2+2a_3)\,X_1^4+\mbox{ circ. perm. }\\[4mm]
+a_1(-4a_1^2+3a_2^2+3a_3^2)\,X_1^2+\mbox{ circ. perm. }\\[4mm] 
+a_1^2(a_1-a_2-a_3)+\mbox{ circ. perm. }+a_1a_2a_3.\earr\eeq
Here the sextic terms in $V_3$ and $I_3$ are again oposite in sign but their quartic terms are {\em different}. Let us recall that the quartic terms in $-V_2$ and $I_2$ (which are equal) were
\[(a_1X_1^2+a_2X_2^2+a_3X_3^2)^2\]
so we cannot express the quartic terms in (\ref{w4}) using such a term. So our integrable potential $V_3$ is indeed different from the potential $I_3$ of Wojciechowski.

It would be quite cumbersome to give a general comparison of the results for the countable set of potentials, but we hope that these arguments are sufficient to show that our integrable potentials are different from the ones considered by Wojciechowski.

\section{Quantization}
Let us discuss briefly the quantization of these models. Since there are 
no quantization ambiguities we do not expect any problem with quantum 
integrability. The quantum observables should verify
\beq\label{qu1}
[\wh{M}_i,\wh{M}_j]=-i\eps_{ijk}\,\wh{M}_k,\qq 
[\wh{M}_i,\wh{X}_j]=-i\eps_{ijk}\,\wh{X}_k,\qq 
[\wh{X}_i,\wh{X}_j]=0,\qq i,j,k=1,2,3.\eeq
For notational convenience we will use also $\wh{M}_i\equiv{\cal Q}(M_i),$ etc... 
Then the classical quantities are unambiguously quantized as
\beq\label{qu2}
\wh{H}_n\equiv{\cal Q}(H_n)=\sum_i a_i\wh{M}_i^2+\wh{U}_n,\qq 
\wh{I}_n\equiv{\cal Q}(I_n)=\sum_i \wh{M}_i^2+\wh{V}_n,\eeq
and the constraints are now operator valued:
\beq\label{qu3}
\wh{C}_1=\sum_i \wh{X}_i^2=Id,\qq \wh{C}_2=\sum_i\wh{X_i}\wh{M}_i=
\sum_i\wh{M}_i\wh{X}_i=0.\eeq
To prove the quantum conservation we start from
\beq\label{qu4}
 [\wh{H}_n,\wh{I}_n]=[\wh{H}^{(2)},\wh{V}_n]-[\wh{I}^{(2)},\wh{U}_n]\eeq 
which gives 
\beq\label{qu5}
[\wh{H}_n,\wh{I}_n]= \sum_i\wh{M}_i\,[\wh{M}_i, a_i\wh{V}_n-\wh{U}_n]+
\sum_i [\wh{M}_i,a_i\wh{V}_n-\wh{U}_n]\,\wh{M}_i.\eeq
One can check that
\beq\label{qu6}  
[\wh{M}_i,a_i\wh{V}_n-\wh{U}_n]=-i{\cal Q}(\{M_i,a_iV_n-U_n\})=
-i{\cal Q}( \wh{L}_i(U_n-a_iV_n)).\eeq
We have seen in (\ref{fin1}) that
\beq\label{qu7} 
\wh{L}_1(U_n-a_1V_n)=\pt_V V_{n-1}
\left(\rule{0mm}{4mm}-2S(X)X_1+2a_1^2(a_2-a_3)(1-C_1(X))X_2X_3\right),\eeq
and since only $X$-dependence is involved one has
\beq\label{qu8}\barr{lcl}\dst 
{\cal Q}(\wh{L}_1(U_n-a_1V_n))& = & \wh{\pt_V V_{n-1}}\left(\rule{0mm}{4mm}
-2\wh{S(X)}\wh{X}_1+2a_1^2(a_2-a_3)\wh{X}_2\wh{X}_3(Id-\wh{C}_1)\right)\\[4mm]\dst 
 & = & -2\wh{\pt_V V_{n-1}}\wh{S(X)}\wh{X}_1.\earr
\eeq
and then replacing this result into (\ref{qu5}) we end up with
\[
[\wh{H}_n,\wh{I}_n]= -2\left(\sum_i\wh{M}_i\wh{X}_i\right)\wh{S(X)}\wh{\pt_V V_{n-1}}
-2\wh{\pt_V V_{n-1}}\wh{S(X)}\left(\sum_i\wh{X_i}\wh{M}_i\right)=0.
\]

This argument is quite rough: it would be more complete if one were able to use true 
coordinates in $T^*\,S^2$ and to quantize the unconstrained theory, using for 
instance the ``minimal" quantization scheme as developed in \cite{DV}. Notice that  
the Lax pair is known for the Neumann system \cite{Su}, but not for these Neumann-like 
systems. This lacking piece of knowledge could possibly be of great help in 
finding the separation variables and handling the unconstrained quantization 
problem mentioned above.

\vspace{2mm}
\noindent{\bf Acknowledgment:} we thank the Referee for pointing out the reference \cite{Wo}.

\end{document}